\documentclass[conference]{IEEEtran}
\IEEEoverridecommandlockouts

\makeatletter
\def\ps@IEEEtitlepagestyle{%
  \def\@oddfoot{\mycopyrightnotice}%
  \def\@oddhead{\hbox{}\@IEEEheaderstyle\leftmark\hfil\thepage}\relax
  \def\@evenhead{\@IEEEheaderstyle\thepage\hfil\leftmark\hbox{}}\relax
  \def\@evenfoot{}%
}
\def\mycopyrightnotice{%
  \begin{minipage}{\textwidth}
  \centering \scriptsize
  Copyright~\copyright~2024 IEEE. Personal use of this material is permitted. Permission from IEEE must be obtained for all other uses, in any current or future media, including\\reprinting/republishing this material for advertising or promotional purposes, creating new collective works, for resale or redistribution to servers or lists, or reuse of any copyrighted component of this work in other works by sending a request to pubs-permissions@ieee.org.
  \end{minipage}
}
\makeatother

\usepackage{cite}
\usepackage{amsmath,amssymb,amsfonts}
\usepackage{graphicx}
\graphicspath{{./images/}}
\usepackage{textcomp}
\usepackage{xcolor}
\usepackage{multirow}
\usepackage{caption}
\usepackage{tabularx}
\usepackage{subfig}
\usepackage{mathtools}
\usepackage{color}
\usepackage{multicol}
\usepackage{listings} 
\usepackage{algorithm}
\usepackage{algorithmic}
\usepackage{pifont}%
\newcommand{\xmark}{\ding{55}}%

\usepackage{url}

\usepackage{breakurl}
\usepackage[breaklinks]{hyperref}

\def\BibTeX{{\rm B\kern-.05em{\sc i\kern-.025em b}\kern-.08em
    T\kern-.1667em\lower.7ex\hbox{E}\kern-.125emX}}
\begin{document}

\title{A Deep-Learning Technique to Locate Cryptographic Operations in Side-Channel Traces}

\author{\IEEEauthorblockN{Giuseppe Chiari}
\IEEEauthorblockA{\textit{DEIB} \\
\textit{Politecnico di Milano}\\
Milan, Italy \\
giuseppe.chiari@polimi.it}
\and
\IEEEauthorblockN{Davide Galli}
\IEEEauthorblockA{\textit{DEIB} \\
\textit{Politecnico di Milano}\\
Milan, Italy \\
davide.galli@polimi.it}
\and
\IEEEauthorblockN{Francesco Lattari}
\IEEEauthorblockA{\textit{DEIB} \\
\textit{Politecnico di Milano}\\
Milan, Italy \\
francesco.lattari@polimi.it}
\and
\IEEEauthorblockN{Matteo Matteucci}
\IEEEauthorblockA{\textit{DEIB} \\
\textit{Politecnico di Milano}\\
Milan, Italy \\
matteo.matteucci@polimi.it}
\and
\IEEEauthorblockN{Davide Zoni}
\IEEEauthorblockA{\textit{DEIB} \\
\textit{Politecnico di Milano}\\
Milan, Italy \\
davide.zoni@polimi.it}
}

\maketitle

\IEEEpubidadjcol
\begin{abstract}
Side-channel attacks allow extracting secret information from the execution of
cryptographic primitives by correlating the partially known computed data and
the measured side-channel signal. 
However, to set up a successful side-channel attack, the attacker has to perform 
\emph{i)}~the challenging task of locating the time instant in which the target
cryptographic primitive is executed inside a side-channel trace and then
\emph{ii)}~the time-alignment of the measured data on that time instant.
This paper presents a novel deep-learning technique to locate the time
instant in which the target computed cryptographic operations are executed in
the side-channel trace. In contrast to state-of-the-art solutions,  the
proposed methodology works even in the presence of trace deformations
obtained through random delay insertion techniques.
We validated our proposal through a successful attack against a variety of unprotected 
and protected cryptographic primitives that have been executed on an FPGA-implemented
system-on-chip featuring a RISC-V CPU.
\end{abstract}

\begin{IEEEkeywords}
Side-channel analysis, locating of cryptographic operations, deep-learning, computer security.
\end{IEEEkeywords}

\section{Introduction}
\label{sec:introduction}
Side-channel attacks emerged as one of the most critical security threats in
modern cryptography since they allow to breach into mathematically secure
cryptographic algorithms by exploiting weaknesses in their physical
implementation.
To extract the secret key from the target implementation of the cryptographic
operation~(CO), side-channel attacks leverage the dependency between the
data being processed and an observable
environmental signal, i.e., the side-channel signal produced by the computing platform.
In the last two decades, several methods, such as
differential power analysis~(DPA)~\cite{KJJ99}, correlation power
analysis~(CPA)~\cite{BCO04}, template attacks~(TA)~\cite{CRR02},
as well as ML-based solutions~\cite{MPP16, CDP17} have been presented
to maximize the effectiveness of the attack by leveraging the
specific attack conditions.
Notably, all the proposed techniques share two common requirements. First, they
require a large number of executions of the same CO with different inputs.
Second, the attacker needs to locate and align in time all the executions of the
CO in the side-channel trace to feed the attack method of choice.

Considering the security assessment of a cryptographic implementation in a
controlled environment, e.g., a laboratory, the attacker is granted full access to
the target device. Moreover, the use of security evaluation boards, e.g., SASEBO
SAKURA-II~\cite{sakura} and NewAE CW305~\cite{cw305}, makes available the so-called
trigger pins that are meant to ease the alignment in time between the executions
of the CO and the corresponding side-channel signals.

In contrast, the security assessment in real-world scenarios requires the attacker
to perform the challenging task of locating the CO within the side-channel trace
without preliminary knowledge of the location of the COs and without support
of a triggering infrastructure.
We note that certain real-world scenarios still allow rough alignment of the measured
side-channel signal with the execution of the CO either \textit{(i)}~by
leveraging specific logic events happening in the computing platform or
\textit{(ii)}~by using pattern-matching techniques applied to the side-channel
trace to generate so-called virtual triggers~\cite{newAeChipwhispererPRO,icWave}.
Recent contributions,
i.e.,~\cite{BFP22,tches2021_semi_automatic_locating_sca}, consider
the identification of the COs in the side-channel trace in the presence of
lightweight noise due to interrupt service routines and single-core
multi-threading context switches.
However, locating the COs in the side-channel trace when the
computing platform implements effective randomization countermeasures, e.g.,
random delay~\cite{durvaux2012} and dynamic frequency scaling~\cite{HDL+20},
still represents a complex and open challenge.

\smallskip\noindent\textbf{Contributions -}
This work presents a deep-learning approach to locate the COs in the
side-channel trace when the target 
device implements an effective random delay countermeasure.
Our proposal presents three contributions to the state of the art:
\begin{itemize}
    \item We present, to the best of our knowledge, the first deep-learning
        approach to locate the COs within the side-channel power trace in
        presence of a random delay countermeasure. The solution automatically
        locates and aligns the COs within the side-channel trace, thus allowing
        to mount a subsequent successful attack that is experimentally
        demonstrated to be impossible otherwise.

    \item We evaluate the proposed solution considering different cryptographic
        primitives and different settings for the random delay countermeasure,
        also performing the CPA attack to assess the quality of the achieved alignment.
    \item We published the tool under an open-source license,
        including a set of traces to allow reproducibility and further
        research. Tool and dataset are available at~\cite{GitRepo}.
\end{itemize}

The rest of the paper is organized into four sections. Section~\ref{sec:SoA}
discusses the academic and commercial tools to locate the COs within 
side-channel traces. Section~\ref{sec:methodology} presents the proposed
deep-learning approach. Section~\ref{sec:expEval} details the experimental
results. Finally, Section~\ref{sec:conclusions} presents the conclusions.

\section{Related Works}
\label{sec:SoA}
Apart from the use of the so-called trigger signals made available in the
security evaluation boards, e.g., SASEBO SAKURA-II~\cite{sakura}, NewAE
CW305~\cite{cw305}, few solutions exist to locate the COs by
matching a previously computed CO template in the side-channel trace.
From commercial viewpoint, Riscure icWaves~\cite{icWave} and NewAE
ChipWhisperer Pro~\cite{newAeChipwhispererPRO} are two FPGA-based devices
offering virtual-triggering capabilities.
In particular, these devices generate a trigger pulse after the real-time
detection of a pattern in the monitored side-channel trace. 

However, using architectural-level techniques to morph the power trace
represents an easy-to-implement and effective countermeasure to deceive
pattern-matching-based solutions. For example,
\cite{2016multiThreadCounterLocatingCO} presents how time-sharing multithreading
on a single-core microcontroller can hinder the attacker from correctly locating
the COs, also discussing the use of interrupt service routines as a
means to morph the shape of the side-channel trace.
In this scenario, advanced solutions to locate the COs in the side-channel
trace emerged. For instance, \cite{BFP22} presents a
computationally efficient technique based on matched filters to locate the
AES-128 cryptosystem in a power trace. The proposed technique works even in
presence of interrupts that can morph the shape of the COs in the power trace.
\cite{cosade2016_becker} proposes a waveform-matching-based triggering system
that is meant to locate the COs in the side-channel trace by matching a
previously computed template of the CO.
\cite{tches2021_semi_automatic_locating_sca} discusses a technique to locate
the COs in a side-channel trace in the presence of interrupts and without a
previously computed template of the CO.

Nevertheless, the current state-of-the-art techniques cannot locate the
COs in the side-channel trace when the computing platform is protected by
effective morphing countermeasures, e.g., random delay.
To this end, we propose a deep-learning-based technique to locate the COs in
power traces collected from a computing platform that implements the random
delay countermeasure.

\section{Methodology}
\label{sec:methodology}
\begin{figure*}[!t]
    \centerline{\includegraphics[width=0.80\textwidth]{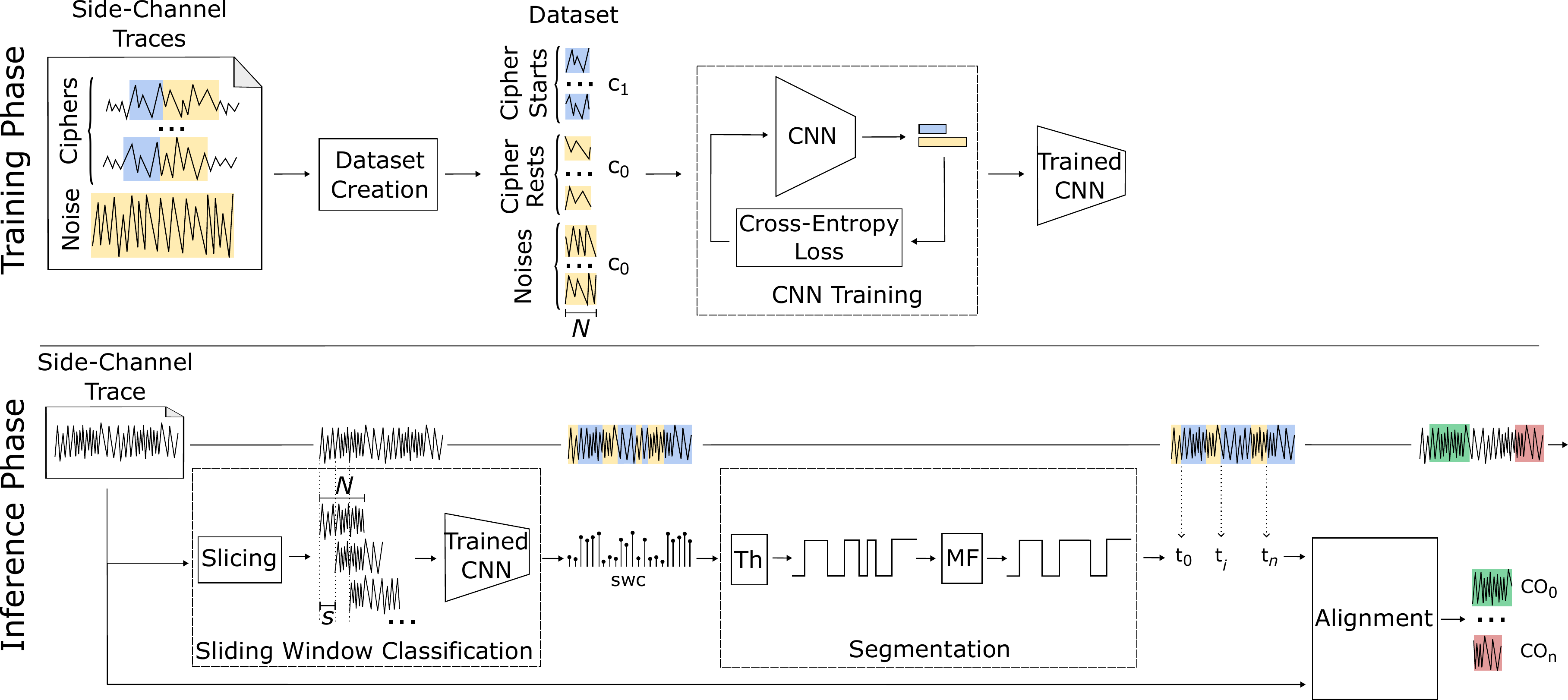}}
    \caption{Overview of the proposed pipeline for locating cryptographic
    operations, divided into training and inference phases. In the Segmentation
    block, \texttt{Th} and \texttt{MF} identify the threshold and median filter
    procedures.}
    \label{fig:pipeline}
\end{figure*}

This section overviews 
the developed deep-learning method for locating
the COs in a side-channel trace when the computing platform implements the
random delay countermeasure to effectively desynchronize the measurements.
The details related to \emph{(i)}~the creation of the dataset,
\emph{(ii)}~the architecture of the proposed Convolutional Neural Network~(CNN)
architecture, as well as \emph{(iii)}~the sliding window classifier, and
\emph{(iv)}~the segmentation procedure are discussed in
Section~\ref{ssec:dataset_meth}, Section~\ref{ssec:cnn_meth},
Section~\ref{ssec:sliding_win_class}, and Section~\ref{ssec:segmentation},
respectively.

\smallskip\noindent\textbf{Threat model -} 
Similarly to profiled attacks, we assume the attacker has access to
an identical copy of the target device, i.e., a clone. However, we target a
real-case scenario in which the attacker can only execute the applications of
choice and measure the corresponding side channel without having complete and
full control over the clone device. In particular, the attacker cannot
activate/deactivate the desynchronization countermeasure, i.e., random delay,
nor can manipulate the hardware by inserting trigger pins.

\smallskip\noindent\textbf{Training and inference pipelines -} 
The proposed deep-learning methodology consists of training and 
inference pipelines~(see Figure~\ref{fig:pipeline}).

The goal of the \emph{training pipeline} is to train a binary classifier that 
allows classifying a slice of $N$ samples from the side-channel trace by labeling 
it either as \emph{beginning of the CO} or not~(see \textit{Training Phase}
in Figure~\ref{fig:pipeline}).
Notably, the methodology is meant to identify the beginning of the COs in
the side-channel trace without removing the desynchronization since, once 
the COs are realigned, the random delay does not represent an effective
countermeasure to state-of-the-art side-channel attacks~\cite{DRS+13}.
We assume the attacker can use an identical copy of the target platform to create
a noise trace and a set of cipher traces to feed to the training pipeline.
The noise trace is a trace obtained from the execution of multiple subsequent
applications different from the CO.
Each cipher trace is collected during the execution of a single CO,
where the attacker can choose the plaintext and the secret key. The desynchronization
mechanism is active when collecting all the traces since the attacker cannot turn it off.
Starting from the collected raw traces, the \emph{Dataset Creation} block assembles
a database that consists of a set of $N$-sample time windows extracted from the collected
noise and cipher traces. For each cipher trace, the first window is labeled as
\emph{beginning of the CO}, while all the remaining windows in the trace and all the windows extracted from the noise trace are labeled as
\emph{not beginning of the CO}. Notably, the number of collected cipher traces,
the length of the noise trace, and the size $N$ of the time windows are configurable parameters.
The dataset is then used to train a CNN binary classifier.

By leveraging the trained CNN, the \emph{inference pipeline} aims at locating the
COs in a new side-channel trace collected from the target device~(see
\emph{Inference Phase} in Figure~\ref{fig:pipeline}).
The inference pipeline consists of three stages: \emph{Sliding Window Classification},
\emph{Segmentation}, and \emph{Alignment}.
The inference pipeline receives a single side-channel trace and outputs its
segmentation to identify the beginning of each CO in the trace. 
At first, the \emph{Slicing} block receives a side-channel trace and
outputs a set of $N$-sample windows that are fed into the CNN classifier.
The sliding amount between two consecutive and partially overlapped windows
represents a configurable parameter of the proposed methodology~(see $s$
parameter in Figure~\ref{fig:pipeline}).
Starting from the classified $N$-sample windows, the \emph{Segmentation}
procedure outputs a vector containing the time instant locating the beginning of
each CO in the side-channel trace.
Finally, the \emph{Alignment} stage cuts the initial side-channel trace
according to the \emph{Segmentation} outputs and aligns the located COs.


\subsection{Dataset Creation}
\label{ssec:dataset_meth}

Considering the training pipeline, the \emph{Dataset Creation} block takes the
side-channel traces in input and generates the dataset to train the CNN~(see
\emph{Training Phase} in Figure~\ref{fig:pipeline}).
The attacker uses the identical copy of the target device to create a noise
trace and a set of cipher traces. The noise trace is obtained from executing
multiple applications different from the CO.
Each cipher trace is collected during the execution of a single CO, where the attacker
can choose the inputs.
Considering the proposed threat model, the attacker
can only execute a chosen application and measure the corresponding
side channel on the clone device without accessing any trigger pins or deactivating
the desynchronization mechanism.
To build the dataset, we replace the unavailable triggering infrastructure with
a set of \emph{NOPs} instructions at the beginning of each CO. The difference in
the power consumption between the execution of the \emph{NOP} instructions and
the execution of the CO allows to easily locate the beginning of the single CO
in each cipher trace.
Notably, the use of the NOPs in the creation of the training dataset is meant to
correctly train the proposed identification infrastructure. 
Once trained, such infrastructure can work on the target architecture that implements
the random delay without inserting any NOP instruction.

For each cipher trace $i$ of length $L_i$ samples, the starting $N$
samples are labeled as \textit{beginning of the CO}~(see $c_1$ class in
Figure~\ref{fig:pipeline}).
The remaining $L_i - N$ samples are equally split into consecutive windows of
width $N$ and labeled as \textit{not beginning of the CO}~(see $c_0$ class in
Figure~\ref{fig:pipeline}).
Moreover, we extract a random set of $N$-sample windows from the noise trace
and we label each of them as \textit{not beginning of the CO}~(see $c_0$ class in
Figure~\ref{fig:pipeline}).


\subsection{Convolutional Neural Network}
\label{ssec:cnn_meth}

\begin{figure*}[!t]
    \centering
    \includegraphics[width=0.85\textwidth]{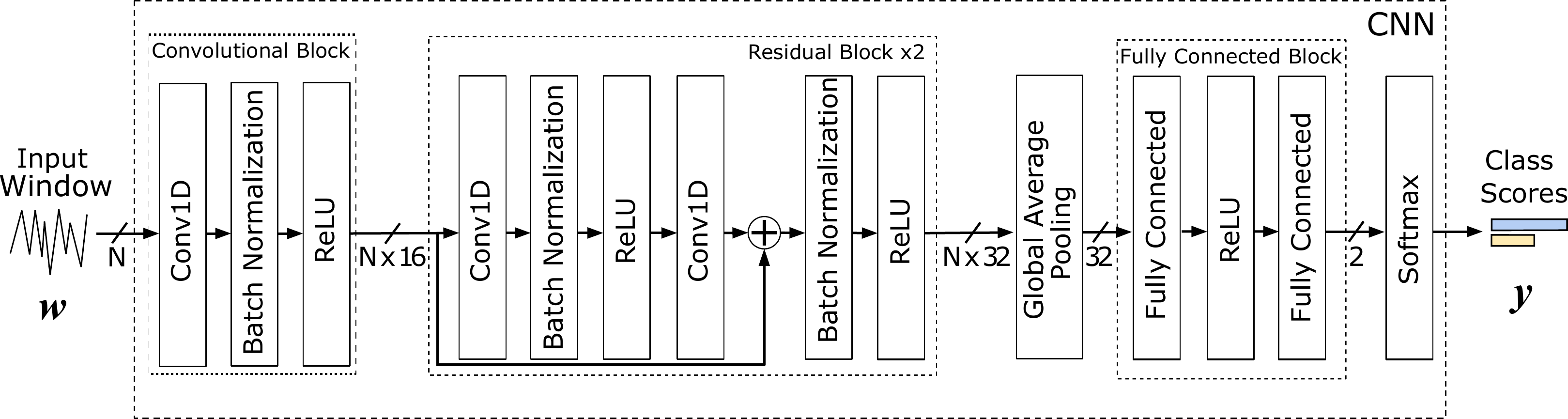}
    \caption{Employed 1D CNN architecture. The network is an adaptation of the known ResNet~\cite{resnet} for 2D image classification.}
    \label{fig:cnn_structure}
\end{figure*}

Figure~\ref{fig:cnn_structure} depicts the architecture of the proposed 1D CNN as
adapted from a 2D ResNet~\cite{resnet}.
The input to the CNN is a window~(\textit{\textbf{w}}) of $N$ samples from a side-channel
trace, while its output is a classification score vector~(\textit{\textbf{y}}).
The CNN architecture comprises six pipelined blocks: a convolutional block,
two residual blocks, a global average pooling layer, a fully connected block,
and a softmax layer.
Each convolutional block features a 1D convolutional layer, a batch normalization
layer~\cite{BN}, and a ReLU activation function~\cite{relu}.
The residual block~\cite{resnet} implements two convolutional blocks enhanced with shortcut
connections to sum the features element-wise~(see
Figure~\ref{fig:cnn_structure}).
Shortcut connections in convolutional blocks are used to improve the training.
All the 1D convolutional layers implement a kernel with size 64, a stride equal
to 1, and zero padding to keep the number of samples equal to $N$.
The first convolutional layer and the one in the first residual block implement
16 filters, while the second residual block increments the number of filters to 32.

The global average pooling layer averages the obtained features over the
temporal dimension $N$, thus reducing the feature vector size from 
$N \times 32$ to $1 \times 32$.
The feature vector is then fed to the two fully-connected layers with a ReLU
activation function.
Finally, the softmax layer outcomes a classification vector with the class scores. 
Notably, the structure of the global
average pooling layer allows the use of different $N$ values for the
training and inference pipelines.

The error between the labels and the output of the network is computed by 
means of the cross-entropy loss function defined in
Equation~\ref{eq:cross_entropy_loss}, where $\textit{\textbf{c}} \in \{0,1\}^2$
is the one-hot encoding of the class label associated to
window~\textit{\textbf{w}}.
\begin{align}
    \label{eq:cross_entropy_loss}
    \mathcal{L}(\textit{\textbf{y}},\textit{\textbf{c}}) = - \sum_{j=1}^{2}
    \textit{\textbf{c}}_j log( \textit{\textbf{y}}_j) = - \sum_{j=1}^{2}
    \textit{\textbf{c}}_j log
    (\textit{g}_j(\textit{\textbf{w}},\boldsymbol{\theta}))
\end{align}
%

\subsection{Sliding Window Classification}
\label{ssec:sliding_win_class}
Considering the \emph{inference pipeline}, the \emph{Sliding Window
Classification} block takes a new side-channel trace, slices it into $N$-sample
windows, and uses the trained CNN to output a classification score to label
each window as \emph{beginning of the CO} or not~(see \emph{Sliding Window
Classification} in Figure~\ref{fig:pipeline}).
The \emph{Slicing} block implements a sliding window procedure to slice the
side-channel trace in input. It takes three inputs, i.e., the
side-channel trace~(\emph{trace}$_{inf}$), the size of the sliding window~($N$),
and the stride~($s$), and outputs an ordered set of $N$-sample windows to feed the
CNN.

The CNN outputs a new signal where each sample is a classification value.
The softmax output of the CNN is a probability distribution of classes, 
which has been observed to hide a structured recurrent pattern that can be 
exploited for locating the COs.
The pattern is more visible in the linear output of the fully-connected block,
which makes localization easier. 
Thus, we consider as inference CNN output the fully-connected score outcome of 
class one.
Depending on whether the $N$-sample window has been classified as the beginning of
the CO or not, a higher or lower class score is assigned to the corresponding windows.
Notably, the CNN can produce a noisy output that cannot be directly used to
infer the precise location of the COs. To this end, the linear output of the CNN
is polished by the subsequent segmentation stage in the inference pipeline~(see
\emph{Segmentation} in Figure~\ref{fig:pipeline}).
%

\subsection{Segmentation}
\label{ssec:segmentation}
The segmentation stage takes the sliding window classification output $swc$ and refines
it to locate the beginning of the COs in the side-channel trace.
It outputs a list of samples that mark the start of each CO.
At first, the algorithm transforms $swc$ into a square wave signal by comparing each sample of
$swc$ with a threshold. For each sample in $swc$, a corresponding output of value -1 or +1
is defined depending if the sample is below or above the threshold value,
respectively~(see \emph{Th} in Figure~\ref{fig:pipeline}).
Then, a median filter~(MF) is applied to further improve the accuracy of the
obtained square wave~(see \emph{MF} in Figure~\ref{fig:pipeline}).
MF is fed with the square wave signal and
a size $k$, which gives the size of the median filter window.
The window slides over the fed square wave signal for which each sample is
replaced with the median value of its $k$ neighbors.
At last, the algorithm returns the number of the samples identifying the rising
edges, i.e., the points in the obtained square wave signal where two consecutive
samples assume a value of -1 and +1, respectively.
The number of the sample is multiplied by $s$,
i.e., the stride value used during the sliding windows classification.
Such points are identified as the beginning of each CO in the analyzed
input trace.

\section{Experimental Evaluation}
\label{sec:expEval}
\begin{figure*}[t]
    \centering
    \subfloat[AES.]{\
		\includegraphics[width=0.15\textwidth]{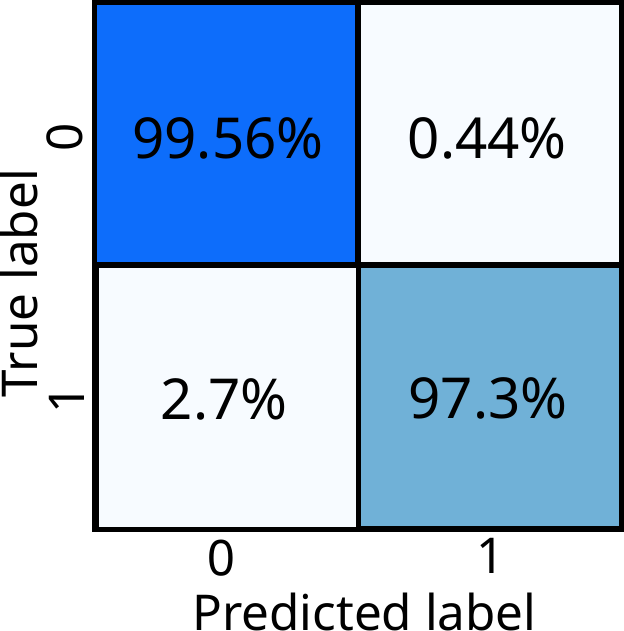}
    \label{sfig:cm_aes}}
    \qquad
    \subfloat[AES mask.]{\
		\includegraphics[width=0.15\textwidth]{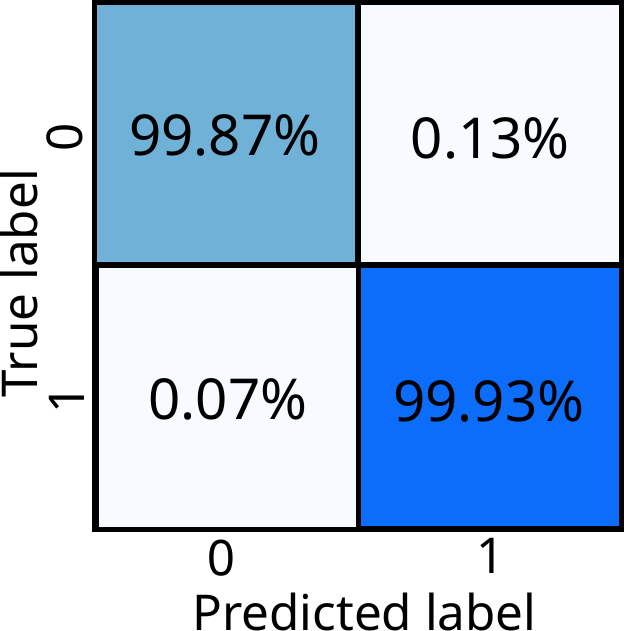}
    \label{sfig:cm_aes_masked}}
    \qquad
    \subfloat[Camellia.]{\
		\includegraphics[width=0.15\textwidth]{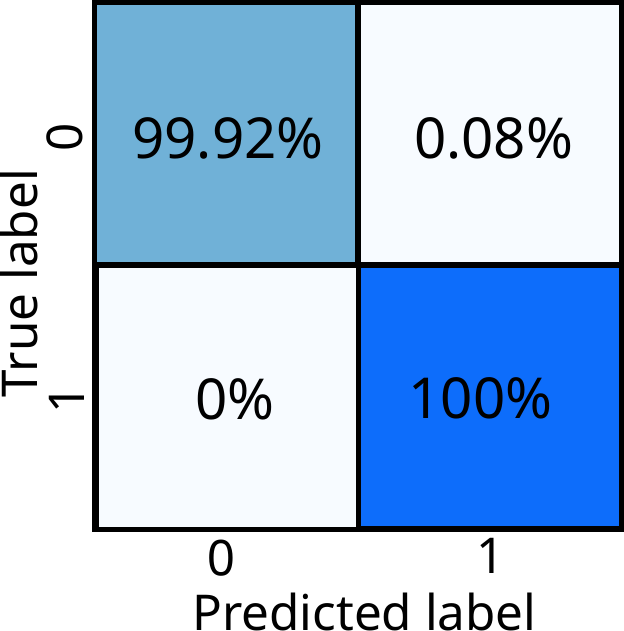}
    \label{sfig:cm_camellia}}
    \qquad
    \subfloat[Clefia.]{\
		\includegraphics[width=0.15\textwidth]{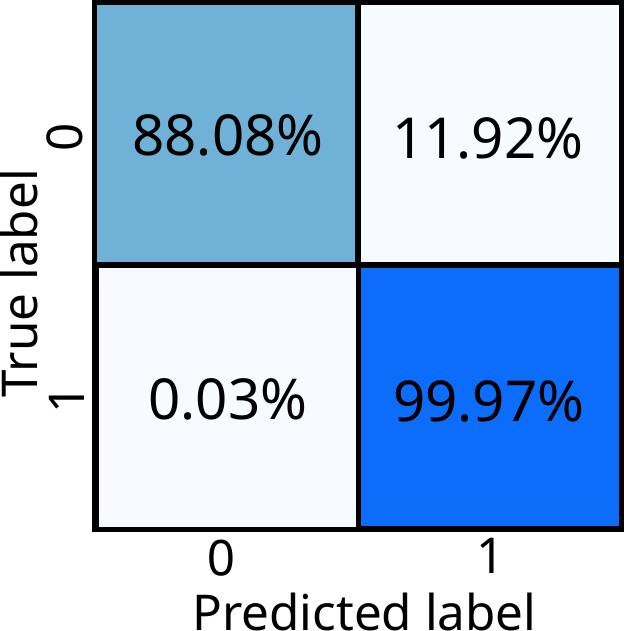}
    \label{sfig:cm_clefia}}
    \qquad
    \subfloat[Simon.]{\
		\includegraphics[width=0.15\textwidth]{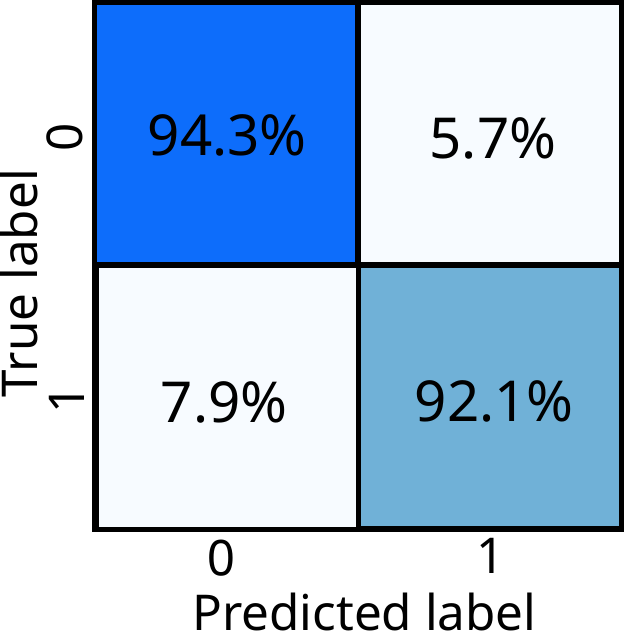}
    \label{sfig:cm_simon}}
    \caption{Test confusion matrices for the different cryptosystems affected by RD-4 random delay.}
    \label{fig:aes_confusion_matrix_test}
\end{figure*}

\begin{table}[t]
	\centering
	\caption{Parameters for each pipeline stage over all the tested ciphers and dataset sizes.
	}
  \scalebox{0.9}{
    \begin{tabular}{c|c|c|c|c|c|c|c}
        \multirow{3}{*}{\textbf{Cipher}}     & \multirow{2}{*}{\textbf{Mean}} 	 & \multicolumn{3}{c|}{\textbf{Pipeline Parameters}} & \multicolumn{3}{c}{\textbf{Dataset Size} (N. Windows)}\\ \cline{3-8}
        & \multirow{2}{*}{\textbf{length}} & \multirow{2}{*}{\textbf{\emph{N}}$_{train}$} & \multirow{2}{*}{\textbf{\emph{N}}$_{inf}$} & \multirow{2}{*}{$\textbf{s}$} & \textbf{Cipher} & \textbf{Cipher} & \multirow{2}{*}{\textbf{Noise}}  \\
        &                 &                                              &                                            &                               & \textbf{Start}  & \textbf{Rest} & \\ \hline\hline
        \textbf{AES}	    & 220k	 & 22k   & 20k & 1k  &  $65\,536$ & $65\,536$ & $32\,768$ \\ \hline
        \textbf{AES\,mask}	    & 50k	 & 4.8k   & 5k & 100  &  $131\,072$ & $65\,536$ & $65\,536$ \\ \hline
        \textbf{Clefia} 	& 108k   & 6k	   & 6k  & 500 &  $65\,536$ & $32\,768$ & $32\,768$ \\ \hline 
        \textbf{Camellia}	& 6k	   & 1.4k  & 1k  & 100 &  $32\,768$ & $65\,536$ &	$32\,768$ \\ \hline 
      \textbf{Simon}      & 10k    & 2k	   & 2k  & 100 &  $65\,536$ & $32\,768$ & $32\,768$ \\ \hline 
    \end{tabular}
    }
	\label{tbl:pipeline_params}
\end{table}

This section discusses the experimental results of the proposed deep-learning
method to locate the execution of COs in a side-channel trace while the
computing platform implements the random delay as an effective trace
desynchronization countermeasure.
The rest of this section is organized into three parts.
Section~\ref{ssec:exp_setup} details the experimental setup.
Section~\ref{ssec:class_seg} presents the results of the CNN training and the
inference pipeline.
Section~\ref{ssec:cpa_eval} discusses a complete example highlighting a
successful side-channel attack, as well as the comparison with two
state-of-the-art proposals.


\subsection{Experimental Setup: Hardware and Software}
\label{ssec:exp_setup}

As validation platform, we chose the NewAE CW305 board~\cite{cw305},
featuring a Xilinx Artix7-100 FPGA.
The power traces were collected with a Picoscope 5244d digital sampling
oscilloscope~(DSO), sampling at 125~Msamples/s with a resolution of 12~bits.
We employed a 32-bit RISC-V System-on-Chip~\cite{lamp} as
reference computing platform deployed on the FPGA and clocked at 50~MHz.
The CPU has been modified to implement the random delay
mechanism at hardware level by leveraging a true random number
generator~(TRNG)~\cite{GGF+22}.
At run-time, the TRNG keeps generating random numbers to determine the actual
number of random instructions to be inserted between each pair of consecutive
program instructions.
The reported results consider two random delay configurations,
i.e., RD-2 and RD-4. RD-2 and RD-4 limit the maximum number of inserted random
instructions between two consecutive instructions in program order to 2 and 4,
respectively.
As the cryptographic operation~(COs) of choice, we selected the constant-time,
unprotected version of four ciphers, i.e., AES-128, Camellia-128, Clefia-128,
and Simon-128, from the OpenSSL software codebase~\cite{openSSL}, and
a masked version of Tiny-AES-128~\cite{medity}.


\subsection{CNN Evaluation}
\label{ssec:class_seg}
This section focuses on the performance evaluation of the CNN.
First, we detailed the dataset creation and the training metrics. Second, we
reported the inference segmentation scores.
A different CNN has been trained with an ad-hoc dataset for each tested cipher.

\smallskip\noindent\textbf{CNN training -} 
The training of the CNN leverages an NVIDIA Titan Xp employing the PyTorch
software framework.
The training datasets have been collected following the procedure of
Section~\ref{ssec:dataset_meth}. 
A brief experimental campaign was carried out to find the right balance
between the three window cases, i.e., \emph{cipher start},
\emph{cipher rest}, and \emph{noise}.
Table~\ref{tbl:pipeline_params} reports the dataset sizes as well as
the windows sizes $N_{train}$.
Windows belonging to the ciphers, i.e.,
\emph{cipher start} and \emph{cipher rest}, are taken balanced between
the key bytes.
As in standard deep learning models, we divided the collected datasets into
training, validation, and testing, respectively 80\%, 15\%, and 5\% of the
total.

Each network was trained for 2 epochs using Adam~\cite{Adam} to minimize
the cross-entropy loss~(see Equation~\ref{eq:cross_entropy_loss}).
The mini-batches size was set to $64$ with a learning rate of $0.001$.
The validation error was evaluated after each epoch, and the network 
with the lowest error was selected.
As an evaluation of the goodness of the trained CNNs, their Confusion Matrices
are shown in Figure~\ref{fig:aes_confusion_matrix_test},
where for each cipher is reported the score on the RD-4 configuration.
The column indices represent the true classes, while the row indices represent
the predicted ones.
Notably, the trained classifiers can discriminate well between
the two classes, as highlighted by the high percentages on the main diagonal of each matrix.

\smallskip\noindent\textbf{Sliding window classification and segmentation -}
The pipeline parameters vary depending on the characteristics of the cipher that
we want to classify, e.g., the average execution time. To this end, we
experimentally determined all the required parameters.
Table~\ref{tbl:pipeline_params} shows the different values of the inference 
window sizes $N_{inf}$ and the strides $s$, for each cipher.
The use of the global average pooling layer allowed a smaller window size
$N_{inf}$ for the inference phase than the one $N_{train}$ used for the training
phase.

We measure the performance of \emph{Sliding Window Classification} and \emph{Segmentation}
blocks by calculating the percentage of \textit{hits}.
This is the ratio of COs correctly located to the total number of true COs
present in the trace.
For each cipher, we tested the inference pipeline considering \emph{i)}~consecutive cipher
executions and \emph{ii)}~encryptions interleaved with random applications. 
The execution of a sequence of consecutive COs assesses
the robustness of the proposed method in locating the COs when they are executed
one after the other. The execution of the COs mixed with noisy applications assesses the
effectiveness of the methodology in locating the COs within a heterogeneous
application scenario. 

The segmentation \textit{hits} score is 100\% for every cryptographic algorithm in both
scenarios, i.e., consecutive encryption and interleaved with noisy applications,
always managing to find all $512$ executions.
The same results are achieved for both random delay configurations, i.e., RD-2 and RD-4. 
This demonstrates the generability of our approach, which, in addition to working
with varying random delay configurations, also works on different encryption
algorithms. 
We also show how our methodology suits protected ciphers, such as masked AES,
whose side-channel traces have great variability.


\subsection{The Complete Attack Flow}
\label{ssec:cpa_eval}
To demonstrate the effectiveness of the proposed locating method,
this section presents a complete attack flow that receives an unknown
side-channel trace and extracts the secret key by leveraging the Correlation
Power Analysis~(CPA) as the effective side-channel attack.
The CPA targets the \emph{sub-byte} intermediate.
A minor aggregation over time is used to fix minor misalignments
due to the rough estimation of the beginning of the COs and to mitigate
the presence of random delay countermeasure.

\begin{table}[t]
	\centering
	\caption{Segmentation and CPA results targeting AES-128. Results
    consider RD-2 and RD-4 settings, 
    and the presence (or not) of noise applications interleaved with the
    COs.
	}
		\begin{tabular}{c|c|c|c|c}
            &\textbf{Random Delay}          & \textbf{Noise} 	  & \multirow{2}{*}{\textbf{Hits} (\%)}    & \textbf{CPA} \\
            &\textbf{Configuration}         &\textbf{Applications}&                                      & (N. COs)  \\ \hline\hline
    \multirow{4}{*}{\textbf{\cite{BFP22}}}  &\multirow{2}{*}{RD-2} & \checkmark	& 0\%  & \xmark  		   \\ \cline{3-5}		
			&				                    & \xmark          & 0\%	 &	\xmark   \\ \cline{2-5}
            &    \multirow{2}{*}{RD-4}          & \checkmark	  & 0\%  &  \xmark   \\ \cline{3-5}		
			&				                    & \xmark          & 0\%	 &  \xmark   \\ \hline
    \multirow{4}{*}{\textbf{\cite{tches2021_semi_automatic_locating_sca}}} &\multirow{2}{*}{RD-2}& \checkmark& 0\%  & \xmark  		   \\ \cline{3-5}		
                                   &                      & \xmark     & 0\%    &	\xmark   \\ \cline{2-5}
                                   &\multirow{2}{*}{RD-4} & \checkmark & 0\%    &  \xmark   \\ \cline{3-5}		
                                   &                      & \xmark     & 0\%	 &  \xmark   \\ \hline
    \multirow{3}{*}{\textbf{This}} &\multirow{2}{*}{RD-2} & \checkmark & 100\%  & 	$3\,695$ \\ \cline{3-5}		
	\multirow{3}{*}{\textbf{Work}} &                      & \xmark     & 100\%  &	$1\,125$  \\ \cline{2-5}
                                   &\multirow{2}{*}{RD-4} & \checkmark	& 100\%  &   $3\,365$  \\ \cline{3-5}		
                                   &				       & \xmark     & 100\%	 &   $1\,220$  \\ \hline
        \end{tabular}
	\label{tbl:AES_results}
\end{table}

Table~\ref{tbl:AES_results} details the attack results considering different
scenarios targeting the AES-128 as the CO of choice. Moreover, we compare against
two state-of-the-art proposals, i.e., \cite{BFP22} and
\cite{tches2021_semi_automatic_locating_sca}. 
For each evaluated scenario, we reported the number of CO executions
in the side-channel trace to achieve a rank equal to 1 in the CPA attacks for
each byte of the secret key. We considered both RD-2 and RD-4 as the
configuration of the random delay mechanism~(see the \emph{Random Delay
Configuration} column in Table~\ref{tbl:AES_results})
We considered the random interleaving of the COs within a set of noisy
applications, as well as the continuous execution of the COs without any noisy
application~(see the \emph{Noise Applications}
column in Table~\ref{tbl:AES_results}).
For each analyzed scenario, our methodology can correctly identify the
beginning of all the COs in the side-channel trace, leading to a successful CPA
attack. In contrast, the two state-of-the-art techniques fail to locate the
COs in the side-channel trace due to the random delay desynchronization,
and thus, the subsequent side-channel attack is unsuccessful.

\section{Conclusions}
\label{sec:conclusions}
We presented a novel technique based on deep learning to identify
cryptographic operations in a side-channel trace. In contrast to
state-of-the-art solutions, the proposed methodology can successfully identify
the COs even when the computing platform implements the random delay mechanism
as the effective desynchronization countermeasure.
We discussed an extensive experimental campaign to locate a series of consecutive
CO executions, as well as the execution of the COs interleaved by other
applications. In particular, the experimental scenarios considered several
cryptographic operations and different implementations of the random delay
mechanisms. The experimental results, extracted from the software execution of
a different set of applications leveraging an FPGA-based RISC-V processor,
confirmed the validity of the proposed methodology that allows to set up a
successful CPA attack also highlighting the limitations of current
state-of-the-art solutions.
To facilitate reproducibility and future research,
we released the tool as open-source software and provided a
collection of test traces.

\bibliographystyle{IEEEtran}
\bibliography{bibliography}

\end{document}